\documentclass[aps,prd,twocolumn,superscriptaddress,showpacs,floatfix,10pt,nofootinbib]{revtex4}
\usepackage{graphicx}
\usepackage{epsfig}
\usepackage{amsmath}
\usepackage{slashed}
\usepackage{units}
\usepackage{eurosym}
\usepackage{array}

\begin{document}

\title{Brane matter, hidden or mirror matter, their various avatars and mixings:\\ many faces of the same physics}

\author{Micha\"{e}l Sarrazin}
\email{michael.sarrazin@fundp.ac.be} \affiliation{Department of Physics, University of Namur (FUNDP), 61 rue de Bruxelles, B-5000 Namur, Belgium}

\author{Fabrice Petit}
\email{f.petit@bcrc.be} \affiliation{BCRC (Member of EMRA), 4 avenue du gouverneur Cornez, B-7000 Mons, Belgium}

\begin{abstract}
Numerous papers deal with the phenomenology related to photon-hidden photon
kinetic mixing and with the effects of a mass mixing on particle-hidden
particle oscillations. In addition, recent papers underline the existence of
a geometrical mixing between branes which would allow a matter swapping
between branes. These approaches and their phenomenologies are reminiscent
of each other but rely on different physical concepts. In the present
paper, we suggest there is no rivalry between these models, which are
probably many faces of the same physics. We discuss some phenomenological
consequences of a global framework.
\end{abstract}


\maketitle

%
%\pacs{11.25.Wx, 11.30.Er, 95.35.+d, 14.20.Dh}
%
%\author{Micha\"{e}l Sarrazin}
%\email{michael.sarrazin@fundp.ac.be} \affiliation{Department of Physics, University of Namur (FUNDP), 61 rue de Bruxelles, B-5000 Namur, Belgium}
%
%\author{Fabrice Petit}
%\email{f.petit@bcrc.be} \affiliation{BCRC (Member of EMRA), 4 avenue du gouverneur Cornez, B-7000 Mons, Belgium}
%

\section{Introduction}

Since a long time, there is a growing interest for physical models related
to hidden matter or hidden worlds. Numerous theoretical and experimental
works result from this line of thought \cite
{1,2,3,4,5,6,7,8,9,9b,10,11,12,13,14,15,16a,16,16b,17,18,19,20,21}. In those
approaches, the standard model of particles splits into a visible sector (to give
the usual ''baryonic'' matter) and one (or many) hidden sector(s) (to give
hidden ''baryonic'' matter). Two approaches are then considered: the hidden
matter could be dissimulated in other braneworlds located somewhere in a
higher dimensional bulk \cite{3,4,5,6,7} or could be sterile particles in
our four-dimensional spacetime \cite{8,9,10,11}. In both cases, it is
expected that many puzzling phenomena could be explained in such a framework.

For instance, in the mirror matter concept, the parity violation in the weak
interaction could be explained by adding a mirror sector to the particle
content. This idea was first evoked by Lee and Yang in 1956 \cite{1}. Thus,
for each left-handed neutrino, a right-handed mirror neutrino restores
parity. Commonly, in the mirror-world scenario, the standard model splits in
two sectors with an opposite parity symmetry breaking \cite{8,9,10,11}.
Usually, it is assumed that particles and mirror particles do not interact
except through gravitation. Then, mirror matter may exist with similar
properties as usual matter but would be undetectable to us through
electromagnetic radiations. Mirror particles are then considered as
plausible candidates for dark matter\footnote{%
Mirror matter is able to self-interact. Though it is usually assumed that
dark matter does not interact with itself, contrary to visible matter, some
authors underlined that dark matter could be self-interacting after all \cite
{14,15}.} by some authors \cite{9,9b,12,13}. In recent years, extensions
of the original idea were suggested which allow couplings between matter
and hidden matter at the quantum level \cite{8,9,9b,10,11}. These couplings
can include photon-mirror photon kinetic mixing \cite{10,11} or
neutrino-mirror neutrino \cite{8} and neutron-mirror neutron \cite{9,9b}
mass mixing. It must be noted that this work is all reminiscent of the
concept of a shadow universe first considered in 1965 by Nishijima and
Saffouri \cite{2}. The idea has also been extended to many coexisting hidden
sectors without necessarily mirror symmetry \cite{4}.

In the other hand, the braneworld idea is considered as a relevant approach
to unify physics \cite{16a,16,16b}. The idea is that our visible Universe is
a three-dimensional space sheet embedded in a hyperspace (called the bulk).
The particles of the standard model are then trapped along such a sheet,
which is called a braneworld. In some work, several braneworlds, invisible
to each other, live in the bulk each one with its own copy of the standard
model. Several issues are concerned by this approach: the hierarchy between
the electroweak and the Planck scales \cite{16}, the cosmic acceleration or
the dark matter origin \cite{3} for instance. Dark matter could be explained
as a hidden baryonic matter localized on another braneworld, provided that
gravitation can spread enough into the bulk. In addition to gravitational
interaction, photon-hidden photon kinetic mixing are expected to allow
matter coupling between branes \cite{5,6,7,10,11}. In recent theoretical
work it has also been shown that a geometrical mixing must occur between
the matter fields of two braneworlds \cite{17,18,19}. As a consequence,
usual matter (related to the standard model) could leap from our braneworld
toward a hidden one, and vice versa \cite{20,21}.

Obviously, all these approaches $-$ related to braneworlds or mirror
symmetry for instance $-$ share a similar phenomenology where a hidden
matter state can oscillate with the visible matter state. If such a kind of
effect would exist, it would then be legitimate to assume that there should exist
a single unified model which could adequately describe this phenomenon,
avoiding the multiplicity of exotic mathematical and physical solutions. In
the present work, we show that such a model exists indeed and that there is
therefore no competition between the previously mentioned approaches. The
geometrical mixing and the matter swapping between branes, the mass mixing
and the particle-mirror(-hidden) particle oscillations, the
photon-mirror(-hidden) photon kinetic mixing are different phenomena which
probably share a common underlying physics.

In section \ref{model}, we will recall the theoretical description of a
two-brane universe and we consider quantum dynamics of a spin$-1/2$ particle
in this setup. Next, in section \ref{pheno} we underline the different
phenomenologies related to the two-brane universe and show that there exists
a unified approach supporting the idea that matter may exist in two states,
each state having a different location in a higher dimensional bulk. The
existence of a swapping mechanism between these two states (which refer here
to the duality matter/hidden matter or matter/mirror matter) follows then
trivially. Finally, in section \ref{compare}, we discuss possible
experimental conditions in the lab-scale for demonstrating with cold
neutrons the existence of these two states.

\section{Fermion dynamics and electrodynamics in a two-brane world}

$\label{model}$

Any Universe with two braneworlds is equivalent to a noncommutative
two-sheeted spacetime $M_4\times Z_2$ when one follows the dynamics of
particles at low energies \cite{17,18}. Let us consider a two-brane Universe
made of two domain walls (which are two kink-like solitons of a scalar field
$\Phi $) on a continuous $M_4\times R_1$ manifold, the relevant Lagrangian
is:
\begin{eqnarray}
\mathcal{L}_{M_4\times R_1} &=&-\frac 1{4G^2}\mathcal{F}_{AB}\mathcal{F}%
^{AB}+\frac 12\left( \partial _A\Phi \right) \left( \partial ^B\Phi \right)
-V(\Phi )  \nonumber  \\
&&+\overline{\Psi }\left( i\Gamma ^A\left( \partial _A+i\mathcal{A}_A\right)
-\lambda \Phi \right) \Psi  \label{L1}
\end{eqnarray}
where $\mathcal{F}_{AB}=\partial _A\mathcal{A}_B-\partial _B\mathcal{A}_A$. $%
\mathcal{A}_A$ is the $U(1)$ bulk gauge field with the coupling constant $G$%
. $\Phi $ is the scalar field. The potential $V(\Phi )$ is assumed to allow
the existence of kink-like solutions, i.e. of domain walls by following the
Rubakov-Shaposhnikov concept \cite{16a}. $\Psi $ is the massless fermionic
matter field. $\Psi $ is coupled to the scalar field $\Phi $ through a
Yukawa coupling term $\lambda \overline{\Psi }\Phi \Psi $ with $\lambda $
the coupling constant. An effective phenomenological discrete two-point
space $Z_2$ can then replace the continuous real extra dimension $R_1$. This
result is obtained from an approach inspired by the construction of
molecular orbitals in quantum chemistry, here extended to fermionic bound
states on branes \cite{17}. At each point along the discrete extra dimension
$Z_2$ there is then a four-dimensional spacetime $M_4$ with its own metric.
Both branes can then be considered as separated by a phenomenological
distance $\delta =1/g.$ $g$ is proportional to an overlap integral of the
fermionic wave functions of each $3$-brane over the extra dimension $R_1$
(see Ref. \cite{17}). In the following, our brane (respectively the hidden
brane) will be conveniently labeled $(+)$ (respectively $(-)$). An effective
$M_4\times Z_2$ effective Lagrangian can then be defined \cite{17}:
\begin{eqnarray}
\mathcal{L}_{M_4\times Z_2} &=&-\frac 1{4e^2}\mathcal{F}_{+\,\mu \nu }%
\mathcal{F}_{+}^{\mu \nu }-\frac 1{4e^2}\mathcal{F}_{-\,\mu \nu }\mathcal{F}%
_{-}^{\mu \nu }  \nonumber  \\
&&-\varepsilon \mathcal{F}_{+\,\mu \nu }\mathcal{F}_{-}^{\mu \nu }+\overline{%
\Psi }\left( {i{\slashed{D}}_A-M}\right) \Psi  \label{L2}
\end{eqnarray}
from which one gets the two-brane Dirac equation \\$\left( {i{\slashed{D}}_A-M}%
\right) \Psi =0$, such that \cite{17}:
\begin{equation}
\left(
\begin{array}{cc}
i\gamma ^\mu (\partial _\mu +iqA_\mu ^{+})-m & ig\gamma ^5-im_r+i\gamma
^5\Upsilon \\
ig\gamma ^5+im_r+i\gamma ^5\overline{\Upsilon } & i\gamma ^\mu (\partial
_\mu +iqA_\mu ^{-})-m
\end{array}
\right) \left(
\begin{array}{c}
\psi _{+} \\
\psi _{-}
\end{array}
\right) =0  \label{Dirac}
\end{equation}
$\psi _{\pm }$ are the wave functions in the branes $(\pm ).$ $A_\mu ^{\pm }$
(respectively $\mathcal{F}_{\pm \,\mu \nu }$) are the electromagnetic
four-potentials (respectively the electromagnetic tensors) in each brane $%
(\pm )$. $e$ and $\varepsilon $ are effective coupling constants. $m$ is the
mass of the bound fermion on a brane. The off-diagonal mass term $m_r$
results from the two-domain-wall Universe \cite{17}. The derivative operator
is then: $D_\mu =\mathbf{1}_{8\times 8}\partial _\mu $ ($\mu =0,1,2,3$) and$%
\ D_5=ig\sigma _2\otimes \mathbf{1}_{4\times 4}$, and the Dirac operator is
defined as ${\slashed{D}=}\Gamma ^ND_N=\Gamma ^\mu D_\mu +\Gamma ^5D_5$
where: $\Gamma ^\mu =\mathbf{1}_{2\times 2}\otimes \gamma ^\mu $\ and\ $%
\Gamma ^5=\sigma _3\otimes \gamma ^5$. $\gamma ^\mu $ and $\gamma ^5=i\gamma
^0\gamma ^1\gamma ^2\gamma ^3$ are the usual Dirac matrices and $\sigma _k$ (%
$k=1,2,3$) the Pauli matrices. Note that Eq. (\ref{Dirac}) is typical from
noncommutative $M_4\times Z_2$ two-sheeted spacetimes \cite{17}.

Concerning the electromagnetic field, it has been proved \cite{17} that the
five-dimensional $U(1)$ bulk gauge field must be substituted by an effective
$U(1)_{+}\otimes U(1)_{-}$ gauge field in the $M_4\times Z_2$ spacetime. $%
U(1)_{+}$ is the gauge group of the photon field localized on our brane,
while $U(1)_{-}$ is that of the photon field located on the hidden brane.
Here, the Dvali-Gabadadze-Shifman mechanism \cite{16b} leads to the gauge
field localization on the branes \cite{17}. As two branes are considered,
the bulk gauge field $\mathcal{A}_A$ splits into $A_\mu ^{\pm }$. The
electromagnetic field \cite{17}
\begin{equation}
\slashed{A}=\left(
\begin{array}{cc}
iq\gamma ^\mu A_\mu ^{+} & \gamma ^5\Upsilon \\
\gamma ^5\overline{\Upsilon } & iq\gamma ^\mu A_\mu ^{-}
\end{array}
\right)  \label{gauge}
\end{equation}
is introduced in the Dirac equation through ${\slashed{D}}_A\rightarrow {%
\slashed{D}}+\slashed{A}$, according to the $U(1)_{+}\otimes U(1)_{-}$ gauge
group. We get \cite{17}
\begin{equation}
\left\{
\begin{array}{c}
\Upsilon =\varphi +\gamma ^5\phi \\
\overline{\Upsilon }=\varphi ^{*}-\gamma ^5\phi ^{*}
\end{array}
\right.  \label{gaugebis}
\end{equation}
where $\varphi $ and $\phi $ are the scalar components of the field $%
\Upsilon $ and $\overline{\Upsilon }=\gamma ^0\Upsilon ^{\dagger }\gamma ^0$.

One underlines that the equivalence between two-brane models and the present
noncommutative two-sheeted spacetime approach is rather general and relies
neither on the domain-wall features nor on the bulk dimensionality \cite{17}.

\newpage

\section{Interpretation and phenomenology}

$\label{pheno}$

From the above two-brane description it is easy to show that the whole
phenomenology of the interactions between particles and hidden particles (or
between matter and mirror matter) can be recovered.

\subsection{Photon-hidden (or mirror) photon kinetic mixing}

$\label{Kinemix}$

If one assumes a Universe made of two branes, it is logical to deal with a $%
U(1)_{+}\otimes U(1)_{-}$ gauge field theory as explained above. Though each
photon fields live in their own brane, they must undergo a kinetic mixing
given by the Lagrangian term (see Eq.(\ref{L2})):
\begin{equation}
\mathcal{L}_k=-\varepsilon \mathcal{F}_{+\,\mu \nu }\mathcal{F}_{-}^{\mu \nu
}  \label{Lem}
\end{equation}
with $\varepsilon $ the coupling strength. In fact, the hidden photon can be
any kind of photon, such as mirror photon \cite{11} or pseudo-photon \cite
{5,6} for instance. The relation between such a coupling and a brane
description was still shown in a stringy context: a $U(1)$\ gauge field on
the hidden brane is coupled to the $U(1)$\ photon field of our brane thanks
to a one-loop process \cite{5,6}.

We are not discussing here the phenomenology of such a coupling which has
been widely studied by other authors \cite{5,6,10,11}. We just note that a
specific mixing between matter and hidden matter can occur through the
photon-hidden photon kinetic mixing. This can be illustrated in a naive way
in the case of the positronium-hidden positronium oscillations for instance
\cite{10}. Indeed, positronium can decay into photons. Since such photons
are coupled to hidden photons, which can be materialized into hidden
positronium, this allows a coupling between positronium and hidden
positronium \cite{10}. Such a second order coupling must occur for neutral
particles only. In the following, we are focusing rather on more original
matter-hidden matter coupling.

\subsection{Mass and geometrical mixing}

$\label{massgeomix}$

Let us focus on the matter-hidden matter swapping. The off-diagonal terms in
the two-brane Dirac equation (\ref{Dirac}) are related to three terms: $%
ig\gamma ^5$, $-im_r$ and $i\gamma ^5\Upsilon $.

$\mathbf{i}.$ The first term $ig\gamma ^5$ is a geometrical mixing. It is
specific to the braneworld formalism \cite{17,18}. Indeed, as explained
above $g$ is proportional to the overlap integral of the extra-dimensional
fermionic wave functions of each $3$-brane over the fifth dimension $R_1$.
The dramatic influence of this term is emphasized in the next section \ref
{lowener}.

$\mathbf{ii}.$ The second term $-im_r$ is a mass mixing term. In the
following, we will demonstrate that its phenomenology is somewhat different
from that arising from geometrical mixing. The difference results from the $%
\gamma ^5$ matrix which is not present in the mass mixing term. Mass mixing
is often considered for neutron-mirror neutron and neutrino-mirror neutrino
couplings \cite{8,9} or for a coupling with hidden sectors \cite{4}. The
present brane approach is fully compatible with all this work. This is
because no restrictive hypothesis was done concerning the exact nature of
the domain walls. If a wall is a kink-soliton which supports left-handed
neutrino, then the second wall can be an antikink-soliton with righ-handed
neutrino \cite{17}. But we can also imagine a domain-wall pair where
particles share the same parity. Note that in the references cited above,
neutral particles are mainly considered. The present model is much more
general and is fully applicable to any spin$-1/2$ standard model particle be
it neutral or charged.

Another point which deserves further attention concerns neutrinos. In the
present model, neutrino-hidden neutrino coupling is only possible through
mass mixing. Indeed, no geometrical mixing is possible for particles without
any magnetic moment (see section \ref{lowener}).

$\mathbf{iii}.$ The third term $i\gamma ^5\Upsilon $ can be considered as an
electromagnetic coupling resulting from the $U(1)_{+}\otimes U(1)_{-}$ gauge
field. Nevertheless, it must not be confused with the coupling resulting
from the photon kinetic mixing. In addition, its existence is closely
related to $ig\gamma ^5$ and $-im_r$. Indeed, if $g=0$ and $m_r\neq 0$ then $%
i\gamma ^5\Upsilon $ reduces to $i\phi $, whereas if $g\neq 0$ and $m_r=0$
it reduces to $i\gamma ^5\varphi $. If $g=0$ and $m_r=0$, the extra term $%
\Upsilon $ is not required. In addition, it can be shown that $\left|
\varphi \right| $ (respectively $\left| \phi \right| $) should present an
amplitude comparable to that of $g$ (respectively $m_r$) \cite{17}. As a
consequence, the two-brane Dirac equation (\ref{Dirac}) can be rewritten in
the more relevant form:
\begin{equation}
\left(
\begin{array}{cc}
i\gamma ^\mu (\partial _\mu +iqA_\mu ^{+})-m & i\widetilde{g}\gamma ^5-i%
\widetilde{m}_r \\
i\widetilde{g}^{*}\gamma ^5+i\widetilde{m}_r^{*} & i\gamma ^\mu (\partial
_\mu +iqA_\mu ^{-})-m
\end{array}
\right) \left(
\begin{array}{c}
\psi _{+} \\
\psi _{-}
\end{array}
\right) =0  \label{Diracmod}
\end{equation}
with
\begin{equation}
\left\{
\begin{array}{c}
\widetilde{g}=g+\varphi \\
\widetilde{m}_r=m_r-\phi
\end{array}
\right.  \label{coupling}
\end{equation}
where we have just replaced the field $\Upsilon $ and the coupling constants
$g$ and $m_r$ by the effective coupling parameters $\widetilde{g}$ and $%
\widetilde{m}_r$. The gauge field term $\Upsilon $ acts as a correction to
the geometrical and mass mixing terms. In addition, without loss of
generality, we will consider now that $\widetilde{g}\approx g$ and $%
\widetilde{m}_r\approx m_r$ since $\left| \varphi \right| $ (respectively $%
\left| \phi \right| $) should not exceed $g$ (respectively $m_r$) as
explained before. This choice allows to further simplify the model.

\newpage

\subsection{Mass and geometrical mixing at non-relativistic energies}

$\label{lowener}$

As shown in previous work \cite{17,18,19}, it is relevant to derive the
non-relativistic limit of the two-brane Dirac equation. Indeed, the
non-relativistic limit leads to a nice and simple Pauli equation where the
two-state structure and its phenomenological consequences are more easily
tackled. This two-state Pauli equation is: $i\hbar \partial _t\Psi =\left\{
\mathbf{H}_0+\mathbf{H}_{cm}+\mathbf{H}_c\right\} \Psi $, with $\mathbf{H}%
_0=diag(\mathbf{H}_{+},\mathbf{H}_{-})$ and $\Psi =\left(
\begin{array}{c}
\psi _{+} \\
\psi _{-}
\end{array}
\right) $. Here, $\psi _{\pm }$ are Pauli spinors. Note that $\mathbf{H}%
_{\pm }$ are the usual four-dimensional Pauli Hamiltonian expressed in each
branes. Moreover, new fundamental coupling terms appear (in natural units)
\cite{17}:
\begin{equation}
\mathbf{H}_c=\left(
\begin{array}{cc}
0 & im_rc^2 \\
-im_rc^2 & 0
\end{array}
\right)  \label{Hc}
\end{equation}
which is simply the mass mixing term, and

\begin{eqnarray}
\mathbf{H}_{cm}=ig\mu \left(
\begin{array}{cc}
0 & -\mathbf{\sigma \cdot }\left\{ \mathbf{A}_{+}-\mathbf{A}_{-}\right\} \\
\mathbf{\sigma \cdot }\left\{ \mathbf{A}_{+}-\mathbf{A}_{-}\right\} & 0
\end{array}
\right)  \label{Hcm}
\end{eqnarray}
where $\mathbf{A}_{\pm }$ are the magnetic vector potentials in the branes $%
(\pm )$ and $\mu $ is the magnetic moment of the particle. Clearly, $\mathbf{%
H}_{cm}$ relates to a mixed geometrical/electromagnetical coupling. The
coupling strength becomes clearly dependent from the magnetic potential in
each branes. The effect of these new terms is discussed hereafter.

\section{Mass mixing versus geometrical mixing}

$\label{compare}$

In the following, we will mainly study the effect of the mass and
geometrical couplings. The second order coupling related to the kinetic
mixing is not explicitly studied since it can be described through a
correction to the mass mixing term.

\subsection{Spontaneous oscillations between visible and hidden sectors}

$\label{sponta}$

Let us first consider a situation where $V_{\pm }$ are the gravitational
fields felt by the particle in each brane and where there is a magnetic
field $\mathbf{B}_{+}\mathbf{=B}$ in our brane. We assume that $\mathbf{B}%
_{-}\mathbf{=0}$ in the second brane: the existence of hidden magnetic
fields will be discussed in section \ref{magfield}. We set $\mathbf{A}=%
\mathbf{A}_{+}-\mathbf{A}_{-}$, the ambient magnetic vector potential such
that $\mathbf{\nabla }\times \mathbf{A\approx 0}$ (see section \ref{vecpot}%
). In addition, we assume that $\mathbf{A}\gg \mathbf{A}_0$ where $\mathbf{%
\nabla }\times \mathbf{A}_0=\mathbf{B}$. For the sake of simplicity, we will
consider separately the role of $\mathbf{H}_c$ and $\mathbf{H}_{cm}$.
Therefore, we have to deal with the Hamiltonians:

\begin{equation}
\mathbf{H}_{gm}=\left(
\begin{array}{cc}
V_{+}+\mu \mathbf{\sigma \cdot B} & -ig\mu \mathbf{\sigma \cdot A} \\
ig\mu \mathbf{\sigma \cdot A} & V_{-}
\end{array}
\right)  \label{Pauligeo}
\end{equation}
for the geometrical/electromagnetic mixing, and
\begin{equation}
\mathbf{H}_{mm}=\left(
\begin{array}{cc}
V_{+}+\mu \mathbf{\sigma \cdot B} & im_rc^2 \\
-im_rc^2 & V_{-}
\end{array}
\right)  \label{Paulimass}
\end{equation}
for the mass mixing.

Let us set $\eta =(V_{+}-V_{-})/\hbar $, $b=\mu B/\hbar $ and consider a
particle located initially ($t=0$) on our brane with a polarization state $%
\mathcal{P}=(N_{\uparrow }-N_{\downarrow })/(N_{\uparrow }+N_{\downarrow })$
(with $N_{\uparrow }+N_{\downarrow }=1$). $N_{\uparrow }$ (respectively $%
N_{\downarrow }$) is the probability to find the particle in a spin-up state
(respectively in a spin-down state) at $t=0$ in our brane. At time $t$, the
probability $P_{+,\uparrow }$ (respectively $P_{+,\downarrow }$) to detect
the particle in the up state in our brane (respectively in the down state in
our brane) is given by:
\begin{eqnarray}
P_{+,\uparrow } &=&N_{\uparrow }\left( 1-\frac{4\Omega ^2}{4\Omega ^2+(\eta
-b)^2}\times \right.  \label{Osci1} \\
&&\left. \sin ^2\left( (1/2)\sqrt{4\Omega ^2+(\eta -b)^2}t\right) \right)
\nonumber
\end{eqnarray}
and
\begin{eqnarray}
P_{+,\downarrow } &=&N_{\downarrow }\left( 1-\frac{4\Omega ^2}{4\Omega
^2+(\eta +b)^2}\times \right.  \label{Osci2} \\
&&\left. \sin ^2\left( (1/2)\sqrt{4\Omega ^2+(\eta +b)^2}t\right) \right)
\nonumber
\end{eqnarray}
where $\Omega =\left| m_c\right| c^2/\hbar $ if we consider the mass mixing
or $\Omega =g\mu A/\hbar $ if we consider the geometrical mixing. Eqs. (\ref
{Osci1}) and (\ref{Osci2}) show that the particle undergoes Rabi-like
oscillations between its two states, i.e. between the two branes. It is
important to note that the oscillations are strongly suppressed when $\left|
\eta \right| $ becomes greater than $\Omega $, i.e. when the particle is
strongly interacting with its environment. The most striking point with this
example, is that there is no difference between mass mixing and geometrical
mixing for spontaneous oscillations. The differences will occur in the
interpretation of the experimental results. In the following, we discuss the
values of the unknown parameters of the problem: $g$, $m_r$, $\mathbf{A}$, $%
\eta$, and $\mathbf{B}_{-}$.

\subsection{Magnitude of mass mixing versus geometrical coupling}

$\label{compmag}$

In a previous work \cite{17}, it was shown that $g$ and $m_r$ are related to
different overlap integrals of the extra-dimensional fermionic wave
functions over the extra dimension (see Eqs. (44) and Appendix B in Ref.
\cite{17}). Though the physical interpretation of each integral is different
from each other \cite{17}, it is possible to specify some constraints. For a
brane and a mirror brane, $g$ and $m_r$ have a similar magnitude \cite{17}:
\begin{equation}
g\approx m_r\propto (1/\xi )\exp (-d/\xi )  \label{relagm}
\end{equation}
where $d$ is the distance between each brane and $\xi $ is the brane
thickness (with $\hbar =c=1$). In this case, it is trivial to check from
Eqs. (\ref{Hc}) and (\ref{Hcm}) that when $A>A_c=4mc/eg_s$ (where $m$ is the
mass of the particle, and $g_s$ its Land\'{e} factor), then the effect of
the geometrical mixing becomes larger than that of the mass mixing. This
critical value is $A_c=3.4\cdot 10^{-3}$ T$\cdot $m for the electron, while
it is $A_c=3.3$ T$\cdot $m for the neutron.

These values constitute an important indication for the model. Indeed,
values of the order $10^9$ T$\cdot $m are expected for astrophysical
magnetic potentials (see section \ref{vecpot}). In that case, the
geometrical/electromagnetic coupling $\mathbf{H}_{cm}$ would be larger than
the mass mixing $\mathbf{H}_c$ by nine orders of magnitude. Of course, we
cannot fully reject the possibility of a domain-wall pair such that $g=0$
and $m_r\neq 0$, but there is absolutely no evidence for this to date.

\subsection{Magnitude of ambient magnetic vector potentials}

$\label{vecpot}$

Following our previous remark, let us now consider the exact influence that
an astrophysical magnetic potential may have on the particle dynamics. It
must be stressed that the level of magnitude of such potentials has recently
been discussed in the literature \cite{22,23} in order to give a constraint
to the photon mass (which is still assumed massless in the present work).
Let $\mathbf{A}$ be the sum of the potentials of the astrophysical objects
(galaxy, star, planet) surrounding us. Since each astrophysical object is
endowed with a magnetic moment $\mathbf{m}$, it produces a potential $%
\mathbf{A}(\mathbf{r})=(\mu _0/4\pi )(\mathbf{m}\times \mathbf{r)/}r^3$
corresponding to a magnetic field $\mathbf{B}(\mathbf{r})=\mathbf{\nabla }%
\times \mathbf{A}(\mathbf{r})$.

Following Eq. (\ref{Hcm}), we note that it is the difference $\mathbf{A}=%
\mathbf{A}_{+}-\mathbf{A}_{-}$ between the magnetic potentials of each world
which is important. As $\mathbf{A}_{-}$ depends on hidden sources in the
other braneworld, we cannot set its value. We should then consider $\mathbf{A%
}$ as an unknown parameter. Nevertheless, unless that $\mathbf{A}_{+}$ and $%
\mathbf{A}_{-}$ are fortuitously anti-collinear and almost equal, which is a
very unlikely situation, $\mathbf{A}_{-}$ cannot significantly change the
magnitude of $\mathbf{A}_{+}$. As a consequence, an estimation of $\mathbf{A}%
_{+}$ should be a good approximation of the magnitude of $\mathbf{A}$.

It can be noticed that in the neighborhood of Earth, and at large distances
from sources, $\mathbf{A}$ is almost constant (i.e. $\mathbf{\nabla }\times
\mathbf{A}\approx \mathbf{0}$) and cannot be turned off with magnetic
shields. Since $A\sim RB$ ($R$ is the distance from the astrophysical
source) the magnitude of the contributions to $\mathbf{A}$ can be deduced.
Considering the galactic magnetic field \textbf{(}$B\approx 1$\textbf{\ }$\mu $%
G \textbf{) }in relation to the Milky Way core \textbf{(}$R\approx 1.9\times
10^{19}$\textbf{\ }m\textbf{)} then $A\approx 2\times 10^9$\ T$\cdot $m \cite
{23,24}. When one considers the Coma galactic cluster, values around $A\approx
10^{12}$ T$\cdot $m are also suggested \cite{22,24}. Nevertheless, some
authors consider this last value as irrelevant as a consequence of the
uncertainties on the magnitude of the magnetic fields inhomogeneities at
extragalactic scales \cite{23}. The order of magnitude $A=10^9$\ T$\cdot $m
is then usually considered as reliable. The Earth and Sun contributions ($%
200 $ T$\cdot $m and $10$ T$\cdot $m respectively) can be neglected \cite{21}.

\subsection{Magnitude of gravitational fields in each brane}

$\label{grav}$

In the present context, a crucial issue concerns the values of the
gravitational fields $V_{\pm }$ felt by the particle in each brane and more
specifically the value of $\left| \eta \right| =|V_{+}-V_{-}|/\hbar $.

Although gravitation is expected to spread into the bulk, its stretching
along the extra dimensions is probably limited \cite{16b,25}. Hence, it can
be assumed that the gravitational effects exerted by a mass of a given brane
on masses located in the other brane depend on the distance between both
branes. Typically, any mass $M$ in the hidden brane acts approximately as a
mass $M^{\prime }$ in our brane such that \cite{25}:
\begin{equation}
M^{\prime }\sim M\exp (-kd)  \label{screen}
\end{equation}
where $d$ is the distance between branes, and $k=2\sigma /3\mathcal{M}^3$
with $\sigma $ the brane tension, and $\mathcal{M}$ the bulk Planck mass
\cite{25}. Then, the mass $M$ does not act on our visible world as in its
own brane: the gravitation felt by the particle is different in each brane
and one gets $\eta \neq 0$ unless that $d=0$. As a consequence, in previous
work \cite{17,18,19,20,21}, it was fairly considered that the gravitational
fields of each brane are independent and so $\eta \neq 0$. Contrarily, if
both branes are close enough such that they share almost the same
gravitational field, then $\eta \approx 0$. But since $g$ and $m_r$ also
depend on the brane distance in an exponential way (see Eq.\ref{relagm}),
for very close branes in the limit $d\rightarrow 0$, $g$ and $m_r$ should
then take unrealistic high values. As a consequence, it is hard to believe
that there is no gravitational role (i.e. that $\eta =0$) in the amplitude
of the matter-hidden matter oscillations.

Of course, the value of $\eta \hbar $ is difficult to determine as the
gravitational contribution from the hidden brane ($V_{-}$) is unknown.
However, rough estimates for gravitational potential energy of neutron give $%
V_{+}$ of the order of $500$ eV for the Milky Way core, while the Sun, the
Earth, and the Moon have contributions of about $9$ eV, $0.65$ eV, and $0.1$ meV
\cite{19,20,21}. As a consequence, it can be assumed that the value of $%
\left| \eta \right| \hbar $ should range between a few meV up to a few keV.
At last, it is clear that $\eta \hbar $ is probably time-dependent. This
mainly comes from the Earth motion around the Sun ($\Delta \eta \hbar \sim
0.31$ eV for a neutron on one year) \cite{19,20,21}. Indeed, it seems
unlikely that Earth can be ''close'' enough to a stellar-like mass
distribution (hidden in the other brane) able to induce a time dependence on
timescales of a year or less.

\subsection{Magnitude of hidden magnetic fields}

$\label{magfield}$

Let us now consider the case of hidden magnetic fields. Magnetic fields
becomes significant if $b\approx \eta $ (see Eqs. (\ref{Osci1}) and (\ref
{Osci2})), i.e. if the gravitational potential energy contributions are
similar to or smaller than the particle energy in a magnetic field. For
instance, a magnetic field about $10^5$ G leads to $b\approx 600$ neV. The
possibility of a gravitational potential energy less or equal to this value
is doubtful according to section \ref{grav}. However, let us first imagine
that the gravitational potentials are similar in both branes (i.e. $\eta =0$%
). In that case, the influence of magnetic fields becomes non-negligible and
the hidden magnetic fields created by the invisible masses (those located on
the other brane) also contribute to the particle dynamics. Obviously, the
hidden magnetic fields can be extrapolated from the dark matter distribution
(which is here implicitly considered as being the matter located on the
other brane). In recent years, dark matter maps have been obtained from its
gravitational effects, on the basis of methods relying on gravitational
lensing or calculations related to the dynamics of visible objects (stars,
nebulae, ...) in interstellar or intergalactic medium \cite
{26,27,28,29,30,31,32}.

Recent observations have confirmed that dark matter constitutes a halo
around the visible Milky Way. In addition, accurate simulations show that
existence of local dense clouds of dark matter can be fairly excluded (the
probability that the solar system is in a dense subhalo is $10^{-4}$ \cite
{29,30}). Therefore, the density of local dark matter cannot exceed that of
the halo which is $\rho _{DM}=4\cdot 10^{-25}$ g$\cdot$cm$^{-3}$ \cite{31}. If one
assumes the existence in our solar system of a hidden molecular cloud made
of hidden H$_2$ (mirror dihydrogen for instance), it should then exhibit a
particle density lower than $0.1$ cm$^{-3}$. Such a density is two orders
lower than the lower densities of known molecular clouds in the visible
Universe \cite{33}. From observations, the strongest magnetic fields in
dense parts of clouds (with densities $\rho $ about $10^4$ up to $10^7$ cm$%
^{-3}$) varies between $0.1$ and $30$ mG \cite{33}. For the most common
parts in clouds (with densities $\rho $ about $10$ up to $10^3$ cm$^{-3}$),
the magnetic fields are about $5$ up to $30$ $\mu $G. On the whole, the
magnetic field in a molecular cloud varies as $\rho ^\alpha $ where $%
0.47\leq \alpha \leq 2/3$ according to the kind of molecular cloud \cite{33}%
. As a consequence, the magnetic field induced by dark matter cannot exceed $%
0.5$ $\mu $G and could be even lower by many orders of magnitude. This value
must be compared with the lowest magnetic fields in current ultracold
neutrons experiments which are around $10$ $\mu $G \cite{34,35}. Then, when
the magnetic field is switched off in these experiments, the hidden magnetic
fields still remain negligible in comparison with the residual magnetic
fields in the neutron vessel. Finally, if the gravitational potentials are
different in both branes ($\eta \neq 0$), it is expected that $\eta \gg b$.
As a consequence, in all cases, the influence of hidden magnetic fields can
be neglected.

\subsection{Consequences}

$\label{consequence}$

The fact that particle and hidden particle could undergo strictly the same
gravitational influence ($\eta =0$), is the less convincing hypothesis due
to the reasons mentioned in section \ref{grav}. In the following, we admit
that $\eta \neq 0$ and $\Omega \ll \eta $, such that the interactions
between the particle and its environment ensure its confinement in the
brane. Then, the following situations can be considered:

$\mathbf{i.}$ The case $b\approx \eta $ where $b$ is related to ''weak''
magnetic fields ($10^{-1}$ G $<B<10^4$ G). For this level of magnitude, the
neutron magnetic energy is extremely weak if compared with usual
gravitational potentials (a magnetic field of about $10^4$ G corresponds to
an energy of about $60$ neV for a neutron). Clearly, $b\approx \eta $ would
require some kind of fine-tuning of the distance between the branes and
their mass content. Even if this seems rather unlikely, this situation
cannot be completely rejected. Though it is not relevant to consider hidden
magnetic fields (see section \ref{magfield}), it is nevertheless interesting
to assess the neutron disappearance rate against the magnetic field
intensity in our visible world. In the following, it must be kept in mind
that neutrons can undergo elastic or inelastic collisions which will inhibit
any coherent oscillation behavior. From the point of view of a single
particle, each collision resets the oscillatory behavior in a quantum Zeno
like effect. This freezing of the oscillations is expected to increase with
temperature and neutron density. Let us set $b=b_0+\delta b$ such that $%
b_0\equiv \eta $. From Eqs. (\ref{Osci1}) and (\ref{Osci2}), it is clear
that a resonance occurs whenever $b=\eta $, i.e. when $\delta b=0$ (we
consider that $\left| \delta b\right| \leq 2\Omega $). If $\left\langle
t\right\rangle $ is the typical time between two consecutive collisions on
the vessel walls, we get:

\begin{equation}
P_{+,\uparrow }\sim N_{\uparrow }\left( 1-\Omega ^2\left\langle
t\right\rangle ^2\right)   \label{mix1}
\end{equation}
and
\begin{equation}
P_{+,\downarrow }\sim N_{\downarrow }\left( 1-\frac{\Omega ^2}{2b_0^2}\left(
1-\frac{\delta b}{b_0}\right) \right)   \label{mix2}
\end{equation}
assuming that $b\gg \left\langle t\right\rangle ^{-1}$, a condition which is
easily achieved in the present context. For instance, for the lowest
magnetic field here considered ($B=10^{-1}$ G) we get $b\approx 920$ s$^{-1}$
while typically $\left\langle t\right\rangle ^{-1}\approx 10$ s$^{-1}$ \cite{21}%
. Here it is interesting to note that there is an asymmetry in the swapping
rate for different polarization states of the neutron. The swapping should
then lead to a shift of the polarization of a neutron gas. Nearly similar
situations were studied in detail by Berezhiani \textit{et al.} for the
neutron-mirror neutron mass mixing (see Refs. \cite{9,9b} for instance). But
in these cases, no gravitational effects were considered, and the role of $%
\eta $ was played by the neutron magnetic energy due to hypothetical hidden
magnetic fields \cite{9b}. One notes that for a usual statistical
thermodynamical set of neutrons at temperature $T$, one gets: $%
N_{\updownarrow }=(1/2)N_0\text{sech} \left( \hbar b/kT\right) \exp (\mp \hbar
b/kT)$ with $N_0$ the number of neutrons. Then, any measurement should
consider the natural polarization shift to avoid a false positive signal.
For instance, in Ref. \cite{9b} such a thermodynamical constraint was not
explicitly considered and could be a source of fallacy. In return, the
results of Berezhiani and Nesti \cite{9b} could be also a clue of a
geometrical mixing instead a mass mixing. This can be justified by the
magnitude of $\mathbf{H}_{cm}$ likely higher than that of $\mathbf{H}_c$ as
explained in section \ref{compmag}. As the hidden magnetic fields are
negligible (see section \ref{magfield}), the resonance according to the
magnetic field can then be due to the gravitational potentials as shown by
Eqs. (\ref{Osci1}), (\ref{Osci2}), (\ref{mix1}) and (\ref{mix2}).

$\mathbf{ii}.$ The situation $b\ll \eta $ where $b$ is related to ''weak''
magnetic fields ($1$ mG $<B<10^4$ G), is the most probable situation. In
this case, the role of the magnetic field is negligible and there is no
polarization dependence. Due to the expected weakness of $\Omega $ related
to $g$ or $m_r$, we get $4\Omega \ll \eta $. Then the particle would exhibit
oscillations of high frequency and low amplitude between the two worlds. The
probability to observe the particle in the hidden brane can be time-averaged
to:
\begin{equation}
p\sim \frac{2\Omega ^2}{\eta ^2}  \label{proba}
\end{equation}

Considering ultracold neutrons stored in a vessel for instance, they have a
probability $p$ to leak from our world toward the hidden one at each wall
collision. This topic has been discussed in a recent paper \cite{21} where
an upper limit on $p$ has been assessed from experimental values.
Constraints on the parameters of the two-brane world were also specified.

\subsection{Resonant oscillations between visible and hidden sectors}

$\label{resonant}$

The two-brane Pauli equation also supports resonant solutions \cite{17,19,20}
thanks to the geometrical mixing term $\mathbf{H}_{cm}$ (Eq. (\ref{Hcm})).
We may consider for instance, a neutron subject to a rotating magnetic
vector potential \\$\mathbf{A}_p=A_p\left(
\begin{array}{ccc}
\cos \omega t & \sin \omega t & 0
\end{array}
\right) $ (in our brane with an angular frequency $\omega $). We neglect
magnetic fields and still consider the gravitational interaction. The
probability to find the neutron in the hidden brane is then:
\begin{equation}
P(t)=\frac{4\Omega _p^2}{(\eta -\omega )^2+4\Omega _p^2}\sin ^2\left( (1/2)%
\sqrt{(\eta -\omega )^2+4\Omega _p^2}t\right)  \label{Reso}
\end{equation}
where $\Omega _p=g\mu A_p/\hbar $ and $\eta =\left( V_{+}-V_{-}\right)
/\hbar $. When $\omega =\eta $, the particle then resonantly oscillates
between the branes. Such a resonant matter exchange has been investigated
and discussed in recent papers \cite{17,19,20,21}. It was suggested that a
device involving a frequency comb laser source could be a very efficient way
to force the matter swapping between the visible and the hidden sector. In
that case, the intensity of the laser source dictates the efficiency of the
matter swapping rate, which is then potentially unlimited \cite{20,21}.

\section{Conclusion}

We have shown that the geometrical mixing and the matter swapping between
branes, the mass mixing and the particle-hidden particle oscillations, the
photon-hidden photon kinetic mixing are the many faces of the same physics
involving quantum dynamics of particles in a bulk containing several
braneworlds (at least two). Since all these phenomena are deeply
interconnected, any positive result coming from an experiment devoted to one
of them would be a strong signal for the reality of the others. A rich
phenomenology emerges then if one considers hidden matter as matter
localized on a hidden brane. The effects of the geometrical mixing are
probably the most important ones due to the magnitude of the coupling and
because they open the door to an artificial matter exchange between
neighboring branes.

\end{document}